\documentclass[twocolumn]{aastex631}

\received{2025 February 5}
\revised{2025 May 21}
\accepted{2025 May 27}

\submitjournal{ApJL}

\begin{document}

\title{Intermittency in Interplanetary Coronal Mass Ejections Observed by Parker Solar Probe and Solar Orbiter}

\correspondingauthor{Julia Ruohotie}
\email{julia.ruohotie@helsinki.fi}

\author[0000-0002-2999-5457]{Julia Ruohotie}
\affiliation{Department of Physics, University of Helsinki, Helsinki, Finland}

\author[0000-0002-4921-4208]{Simon Good}
\affiliation{Department of Physics, University of Helsinki, Helsinki, Finland}

\author[0000-0001-6868-4152]{Christian Möstl}
\affiliation{Austrian Space Weather Office, GeoSphere Austria, Graz, Austria}

\author[0000-0002-4489-8073]{Emilia Kilpua}
\affiliation{Department of Physics, University of Helsinki, Helsinki, Finland}



\begin{abstract}

Intermittency has been studied extensively in the fast and slow solar winds but to a far lesser extent in interplanetary coronal mass ejections (ICMEs). While ICMEs are often characterized by their relatively smooth, large-scale magnetic flux rope structures, a spectrum of fluctuations is nonetheless present at smaller scales. We have examined kurtosis and its scaling exponents at magnetohydrodynamic inertial scales in 49 ICMEs observed between 0.25 and 1 au by Parker Solar Probe and Solar Orbiter, and compared the results to those obtained for the ICME sheath regions and ambient solar wind intervals. Kurtosis behaves similarly in all intervals studied and presents a universal behavior typical of intermittent time series. The ICMEs displayed a radially invariant level of intermittency, suggesting that they are relatively static, well-developed turbulent environments. In the sheath regions, the level of intermittency increased with distance, indicating that the turbulence is not yet fully developed at small heliocentric distances. In addition to intermittent fluctuations related to turbulence, the sheath regions may possess a population of non-turbulent structures that increase the absolute value of kurtosis.

\end{abstract}



\section{Introduction} \label{sec:intro}

The solar wind structure is modified by magnetohydrodynamic (MHD) turbulence which results in the generation of fluctuations at various scales \citep[e.g.][]{Bruno2013}. In Alfv\'en wave turbulence models, the turbulence is due to nonlinear interactions between counterpropagating Alfv\'en waves which give rise to a transfer of energy across different scales. This energy transfer is evident from the power spectral density (PSD) of the magnetic field fluctuations, which takes the form of a power law with spectral index values usually at around -5/3 or -3/2 in the inertial range \citep[e.g.][]{Chen2016, Verscharen2019}. Nonlinear dynamics in the plasma result in nonlinear energy transfer, giving rise to intermittency, an uneven distribution of energy in the system \citep[e.g.][]{Frisch1995, Sorriso-Valvo1999, Matthaeus2015}. Intermittency increases towards smaller scales leading to nonlinear scaling laws and non-Gaussian statistics which are seen as wider tails in probability density functions (PDFs) of increments. The slow solar wind has been observed to be more intermittent than the fast wind but the radial evolution of intermittency, which shows an increasing trend, has only been observed in the fast wind in the ecliptic plane \citep[e.g.][]{Bruno2019a, Wawrzaszek2021}.

Interplanetary coronal mass ejections \citep[ICMEs; e.g.][]{Kilpua2017a} offer a distinct plasma environment compared to the solar wind. They are often associated with large-scale flux ropes with strong fields, low plasma $\beta$, and rotation over a large angle \citep[e.g.][]{Cane2003}. They are additionally observed to be environments with low cross helicity \citep{Good2020b, Good2022}, a parameter that describes the balance of power between Alfv\'enic fluctuations parallel and anti-parallel to the mean magnetic field. A subset of ICMEs with smooth rotation, low plasma temperature, and relatively small fluctuation amplitudes are called magnetic clouds \citep[MCs; e.g.][]{Burlaga1981}. If an ICME propagates fast enough, a sheath region will form ahead of it \citep[e.g.][]{Kilpua2017a}. These structures consist of the preceding solar wind that has been piled up between the shock and the leading edge of the ICME and they are often characterized by their 'turbulent' nature, i.e. containing large-amplitude fluctuations \citep[e.g.][]{Kilpua2019}. ICME sheaths show a range of dynamic turbulent behavior at 1 au \citep[e.g.][]{Moissard2019, Ruohotie2022, Soljento2023}.

The large-scale dynamics of ICMEs are fairly well known, but the same is not true of their small-scale characteristics. In addition to smoothly varying MCs with relatively simple structures, ICMEs with complex structures and fluctuations at various scales are observed \citep[e.g.][]{Janoo1998, Steed2011}. The observed magnetic structure of ICMEs is affected by large-scale dynamics like the expansion of the ICME and interactions with the surrounding solar wind. Small-scale dynamics giving rise to fluctuations include, for example, the turbulence which has been studied both observationally \citep{Good2023, Ruzmaikin1997} and through simulations \citep{Pezzi2024}. Intermittency has been widely researched in the solar wind and lately also in ICME sheaths \citep{Pitna2021, Kilpua2020, Kilpua2021a}. On the other hand, in ICMEs, intermittency and other mesoscale structures and dynamics have generally gathered less attention, though this has changed in recent years \citep[e.g.][]{Palmerio2024}. Intermittency in ICMEs has previously been studied by \citet{Sorriso-Valvo2021} and \citet{Marquez2023} but to our knowledge no dedicated statistical study has been carried out. The presence of intermittency in ICMEs raises questions about the origin of intermittent fluctuations, how their presence fits with the traditional picture of a smooth flux rope structure, and what effects intermittency may have on the dynamics of the ICMEs.

In this paper, we present a statistical study of magnetic field intermittency in ICMEs observed by Parker Solar Probe \citep[PSP;][]{Fox2016} and Solar Orbiter \citep[SolO;][]{Muller2020} at various heliocentric distances. Intermittency is studied using structure functions, with scale-dependent kurtosis used as the main intermittency measure. The results are compared to those obtained by similar analysis of the preceding and trailing solar wind and in the ICME sheath regions. Additionally, we study the connection between kurtosis and heliocentric distance, speed, spectral index, residual energy, cross helicity, and proton $\beta$.

\section{Data and methods} \label{sec:methods}

In this study, 49 ICMEs observed by PSP and SolO at distances between 0.25 and 1 au are studied. Events are collected from the HELIO4CAST ICMECAT database \citep{Möstl2017, Möstl2020}. Events are required to have good coverage of both magnetic field and plasma data without major data gaps. Recalling that ICMEs are characterized by low $\beta$, only events with average proton $\beta$ of less than 0.5 are included. The minimum duration of ICMEs is set at 3 hours in order to sample most of the inertial range in frequency space \citep[e.g.][]{Kiyani2015}. For one event, the trailing end boundary of the ICME given in the ICMECAT database has been altered via visual inspection due to the possible presence of another structure immediately after the ICME. See the \hyperref[sec:appendix]{Appendix} for the full list of ICMEs used.

In addition to the ICME intervals, three other interval types are examined: ICME sheaths, and upstream and downstream solar wind periods. The minimum duration of sheaths is set to 3 hours, similar to the ICMEs. The durations of the upstream and downstream intervals equal the mean duration of the analyzed ICMEs, namely $\sim11.7$ hours. For upstream and downstream intervals to be included in the study, the ICME cannot be preceded or followed by another large-scale structure (e.g. another ICME or crossings of the heliospheric current sheet) unless the separation is at least 23.4 hours (i.e. twice the mean ICME duration). The magnetic field data from the PSP/FIELDS \citep{Bale2016} and the SolO/MAG \citep{Horbury2020} instruments are used with time resolutions of 0.218 s and $\sim$0.125 s, respectively. The ion moments are from the PSP/SWEAP-SPC \citep{Kasper2016} and the SolO/SWA \citep{Owen2020} instruments, and the time resolution varies between $\sim$1 and 28 s.

Scale-dependent properties of the plasma are often examined using structure functions. By estimating the increments of the magnetic field components $\delta B_i (\tau, t) = B_i (t+\tau) - B_i(t)$ at timescale $\tau$, the structure function of $p$th order for the trace of the magnetic field can be defined as
\begin{equation}
    S^p_B(\tau) = \langle | \delta \textbf{B} (\tau, t)|^p \rangle_t,
\end{equation}
where $|\delta \textbf{B} (\tau, t)| = ( \sum_i \delta B_i^2 ) ^{1/2}$ and angle brackets denote time averaging over the interval analyzed. For the calculation of spectral indices, $\alpha_{\mathrm{PSD}}$, we use the second-order structure function, which corresponds to the PSD \citep{Frisch1995, Dudok2013}.

The main intermittency measure used in this study is kurtosis, 
\begin{equation}
    \kappa(\tau) = \frac{S^4_B(\tau)}{\left( S^2_B(\tau) \right)^2},
\end{equation}
which we derive using the second- and fourth-order structure functions \citep[e.g.][]{Frisch1995, Bruno2013}. Kurtosis quantifies the tailedness of the distribution, with $\kappa = 3$ corresponding to the Gaussian distribution \citep[e.g.][]{Stawarz2016, Roberts2022}. Distributions with fatter tails correspond to kurtosis of higher value and are associated with more intermittent fluctuations. In the case of intermittency, kurtosis shows a power-law scaling $\kappa \sim f^{\alpha_\kappa}$ in the inertial range, with steeper values of the scaling exponent, $\alpha_\kappa$, corresponding to more developed turbulence and a more efficient transfer of energy between scales \citep{Sorriso-Valvo2019, Sorriso-Valvo2021, Telloni2021}. If $\kappa$ is approximately constant with scale, the interval is considered self-similar and thus non-intermittent. When comparing intermittency between different intervals, we adopt a similar approach to that of \citet{Sorriso-Valvo2018}, \citet{Telloni2021}, and \citet{Sioulas2022a} and use $\alpha_\kappa$ as the measure of intermittency. 

The connection between intermittency and normalized residual energy and cross helicity is also studied. Residual energy describes the equipartition of power between magnetic field and velocity fluctuations. For deriving the cross helicity, we use Elsasser variables \citep{Elsasser1950}, $\textbf{z}^\pm = \textbf{v} \pm \textbf{b}$, where $\textbf{b} = \textbf{B}/\sqrt{\mu_0 \rho_{\mathrm{p}}}$. Here, we define the normalized residual energy, $\sigma_{\mathrm{r}}$, and cross helicity, $\sigma_{\mathrm{c}}$, using structure functions \citep[e.g.][]{Sorriso-Valvo2021, Marquez2023},
\begin{equation}
    \sigma_{\mathrm{r}} = \frac{S^2_v - S^2_b}{S^2_v + S^2_b}
\end{equation}
\begin{equation}
    \sigma_{\mathrm{c}} = \frac{S^2_{z^+} - S^2_{z^-}}{S^2_{z^+} + S^2_{z^-}}.
\end{equation}

The majority of the analysis is presented in terms of $kd_{\mathrm{i}}$ instead of the spacecraft frequency $f_{\mathrm{sc}}$. This scaling allows for a more accurate comparison of the different spectra in the plasma frame. The connection between $f_{\mathrm{sc}}$ and $kd_{\mathrm{i}}$ comes from Taylor's hypothesis such that $kd_{\mathrm{i}}$ is defined as $kd_{\mathrm{i}} = \frac{2\pi}{e \langle v \rangle}\sqrt{\frac{m_{\mathrm{i}}}{\mu_0 \langle n \rangle}} f_{\mathrm{sc}}$, where $\langle v \rangle$ and $\langle n \rangle$ are time averages over an individual interval \citep[e.g.][]{Good2023}.

Another way to study PSDs is using wavelet analysis \citep{Torrence1998}. The advantage of this method is the retention of the time dependence, allowing the identification of fluctuation power in both time and scale. While not the main analysis method used in this study, we compute the local intermittency measure \citep[LIM;][]{Farge1992}
\begin{equation}
    \mathrm{LIM}(\tau, t) = \frac{|\omega(\tau, t)|^2}{\langle |\omega(\tau, t)|^2 \rangle_t},
\end{equation}
where $|\omega(\tau, t)|^2$ is the wavelet trace PSD. Using LIM, we can identify fluctuations and the regions within different intervals that contribute most to the intermittency. Additionally, we can derive an equivalent kurtosis measure \citep{Meneveau1991a, Meneveau1991b} to the one obtained with the structure functions, namely
\begin{equation}
    \kappa_{\mathrm{LIM}}(\tau) = \langle \mathrm{LIM}^2(\tau, t) \rangle_t.
\end{equation}
Similarly, $\kappa_{\mathrm{LIM}} = 3$ corresponds to the Gaussian distribution. The differences between kurtoses obtained from structure functions and LIM are briefly discussed in Section \ref{sec:results}.

\section{Results} \label{sec:results}

\subsection{An example event} \label{sec:example}

\begin{figure*}[ht!]
\centering
\includegraphics[width=0.8\linewidth]{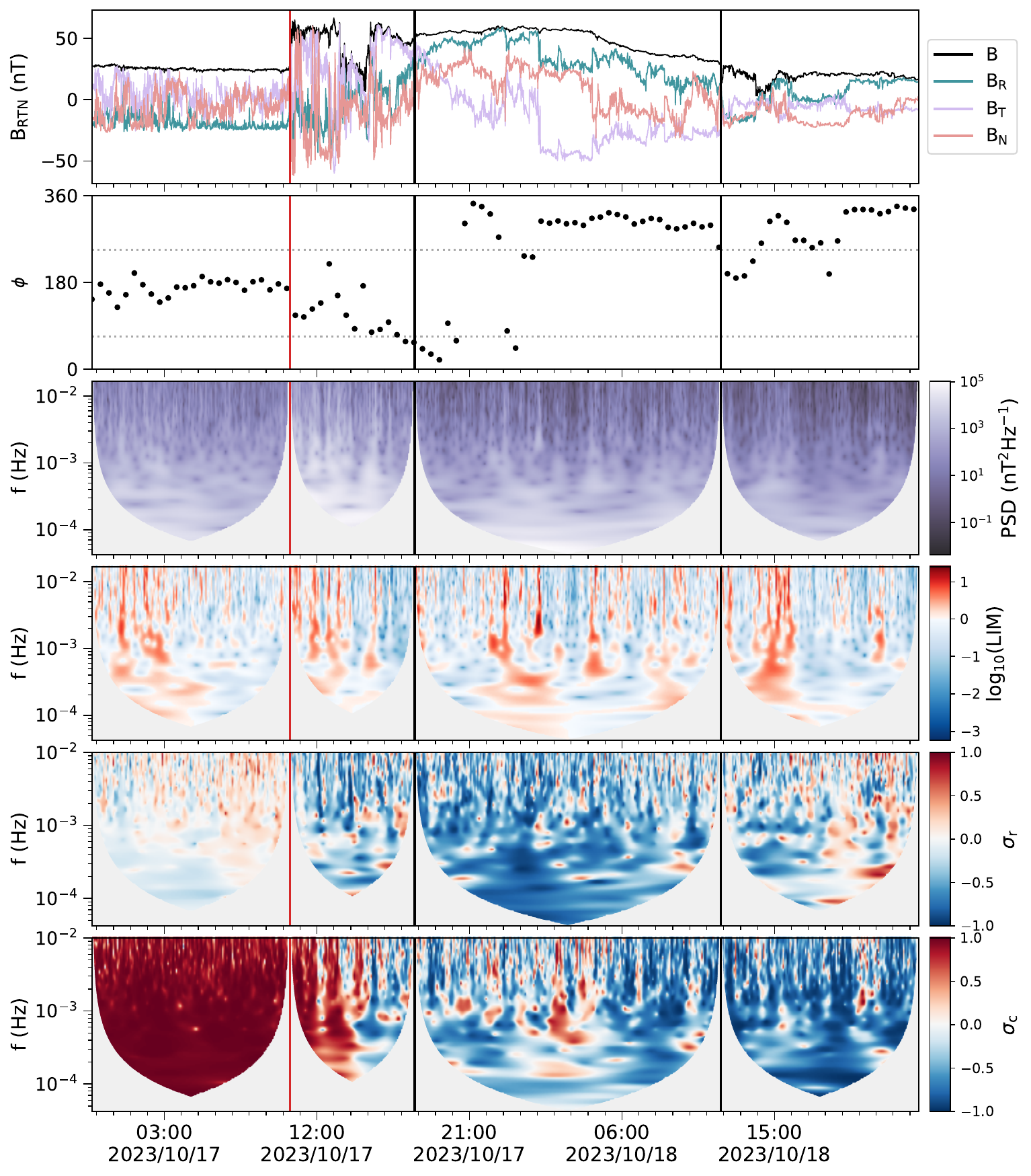}
\caption{An example ICME observed by SolO on October 17, 2023 at 0.37 au. From top to bottom, the panels show magnetic field magnitude and its RTN components, 30-min average of the magnetic field longitude angle in RTN coordinates, wavelet PSD of magnetic field, LIM, $\sigma_{\mathrm{r}}$, and $\sigma_{\mathrm{c}}$. The ICME interval is located between the two vertical black lines.
\label{fig1}}
\end{figure*}

Figure \ref{fig1} shows SolO observations of an ICME observed in October 2023. The first panel shows the magnetic field time series, the second panel shows the 30-min average of the magnetic field longitude angle, and the rest of the panels show wavelet-derived parameters. Before performing the wavelet transform, the time series is divided into separate intervals (upstream, sheath, ICME, downstream) and the transform is applied to the shorter time series separately. Additionally, all normalizations related to calculations of LIM, $\sigma_{\mathrm{r}}$, and $\sigma_{\mathrm{c}}$ are performed separately in these shorter time series. Thus, only parts of the spectrogram falling within the separate cone of influence for each interval are shown in the figure.

From the wavelet PSD we can see that the power in fluctuations increases toward larger scales but some high power fluctuations at smaller scales can also be identified. There is a correspondence between high power fluctuations and high LIM, as one would expect. The colormap of LIM is chosen to highlight regions in red shades with $\mathrm{LIM} > 1$, i.e. times and scales when fluctuations have more power than the average at that scale. High values of LIM are seen across all scales shown in Figure \ref{fig1} and in some cases fluctuations contributing to high LIM values are visible in the magnetic field time series. With this event, both the upstream wind and the sheath have the highest LIM values during the first half of the interval. Within the ICME, high LIM values are seen across the whole interval with the highest observed around 2023-10-17 23:00 and 2023-10-18 01:00 coinciding with sharp fluctuations seen in the magnetic field data. In the downstream interval, the LIM is highlighted around 2023-10-18 15:00 due to enhanced fluctuation amplitude in the magnetic field data. When comparing LIM and $\sigma_{\mathrm{r}}$, there is a clear correlation between high LIM and negative $\sigma_{\mathrm{r}}$. No clear connection between LIM and $\sigma_{\mathrm{c}}$ is seen for this event.

Some systematic inhomogeneities in the distribution of enhanced fluctuation amplitudes across the subintervals may be enhancing the LIM, rather than these enhancements representing intermittency of homogeneous turbulence. In general with the events in this study, the fluctuation amplitude is seen to increase near the boundary regions between different intervals, especially before the leading edge of the ICME in the case the ICME is not accompanied by a sheath. In these cases, the cause is likely the large-scale dynamics between different plasma environments. Due to the possibility of boundary regions affecting the results, we remove 10\% of the interval duration at the end of the upstream and at the start of the downstream, as well as both at the start and the end of the ICME intervals to mitigate the effects of the boundary regions. In the case of sheaths, fluctuations enhancements appear to be much more randomly and variably distributed through the intervals, and so we do not remove any boundary regions from them.

\begin{figure*}[ht!]
\centering
\includegraphics[width=\linewidth]{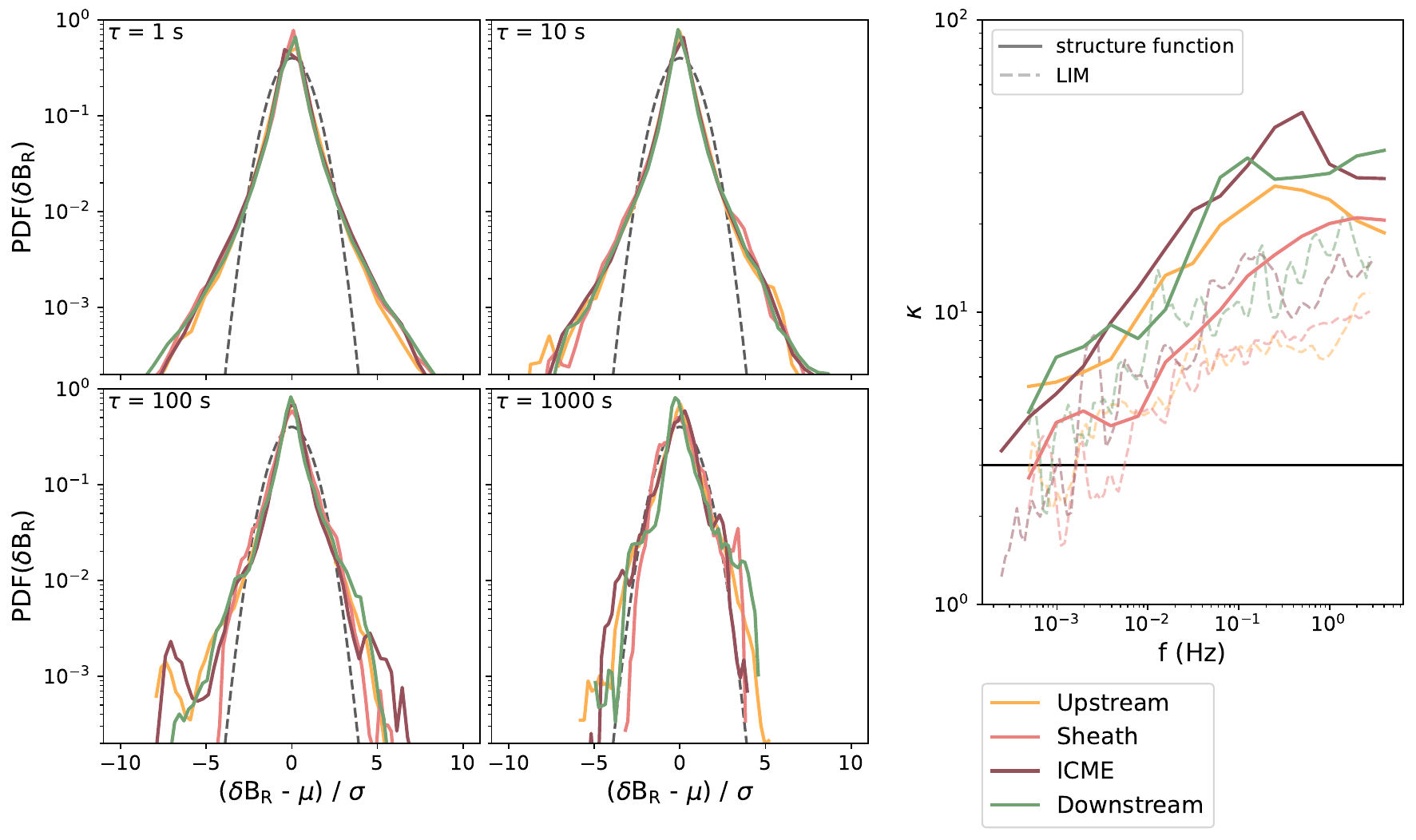}
\caption{The four panels on the left show distributions of $\delta \mathrm{B}_{\mathrm{R}}$ for the example ICME event at different scales. Different colored lines correspond to different intervals (upstream, sheath, ICME, downstream) and dashed line corresponds to a Gaussian distribution. The right panel shows $\kappa$ and $\kappa_{\mathrm{LIM}}$ (dashed lines) as a function of spacecraft frequencies. The black horizontal line corresponds to $\kappa = 3$.}
\label{fig2}
\end{figure*}

The left panels in Figure \ref{fig2} show the distributions of $\delta \mathrm{B}_{\mathrm{R}}$ for the same event as in Figure \ref{fig1}. The distributions are made for each interval separately, differentiated by color. As is characteristic of intermittency, the distributions are rather non-Gaussian at small scales and become more Gaussian as the scale increases. The right panel in Figure \ref{fig2} shows the kurtosis for each interval. All four intervals have a similar trend, where kurtosis is high at small scales and decreases when moving to larger scales (smaller spacecraft-frame frequencies), but the kurtosis of the ICME decreases more steeply, verified by the largest $\alpha_\kappa$ value of 0.40 in the frequency range $10^{-3} - 10^{-1}$ Hz. The scaling exponent values for the upstream, sheath, and downstream are 0.35, 0.28, and 0.36, respectively. Thus, the ICME interval is considered to be the most intermittent for this event. The behaviors of PDFs of $\delta \mathrm{B}_{\mathrm{T}}$ and $\delta \mathrm{B}_{\mathrm{N}}$ (not shown here) are similar to that of $\delta \mathrm{B}_{\mathrm{R}}$ but some differences are seen in the intermittency levels of the components. In the case of $\mathrm{B}_{\mathrm{T}}$, the most intermittent interval is the sheath and the least intermittent is the downstream while for $\mathrm{B}_{\mathrm{N}}$ the situation is opposite: the most intermittent interval is the downstream and the least intermittent the sheath.

The right panel in Figure \ref{fig2} also includes the equivalent kurtosis measure calculated from the LIM (dashed lines). While there is an off-set between the values of $\kappa$ and $\kappa_{\mathrm{LIM}}$, the trend is consistent in all intervals. The scaling exponents of $\kappa_{\mathrm{LIM}}$ for $\delta \mathrm{B}_{\mathrm{R}}$ in the upstream, sheath, ICME, and downstream are 0.22, 0.32, 0.29, and 0.26, respectively. Now the most intermittent interval would be the sheath with the ICME being the second most intermittent interval.

\subsection{Statistical results} \label{sec:statistical}

\begin{figure*}[ht!]
\centering
\includegraphics[width=\linewidth]{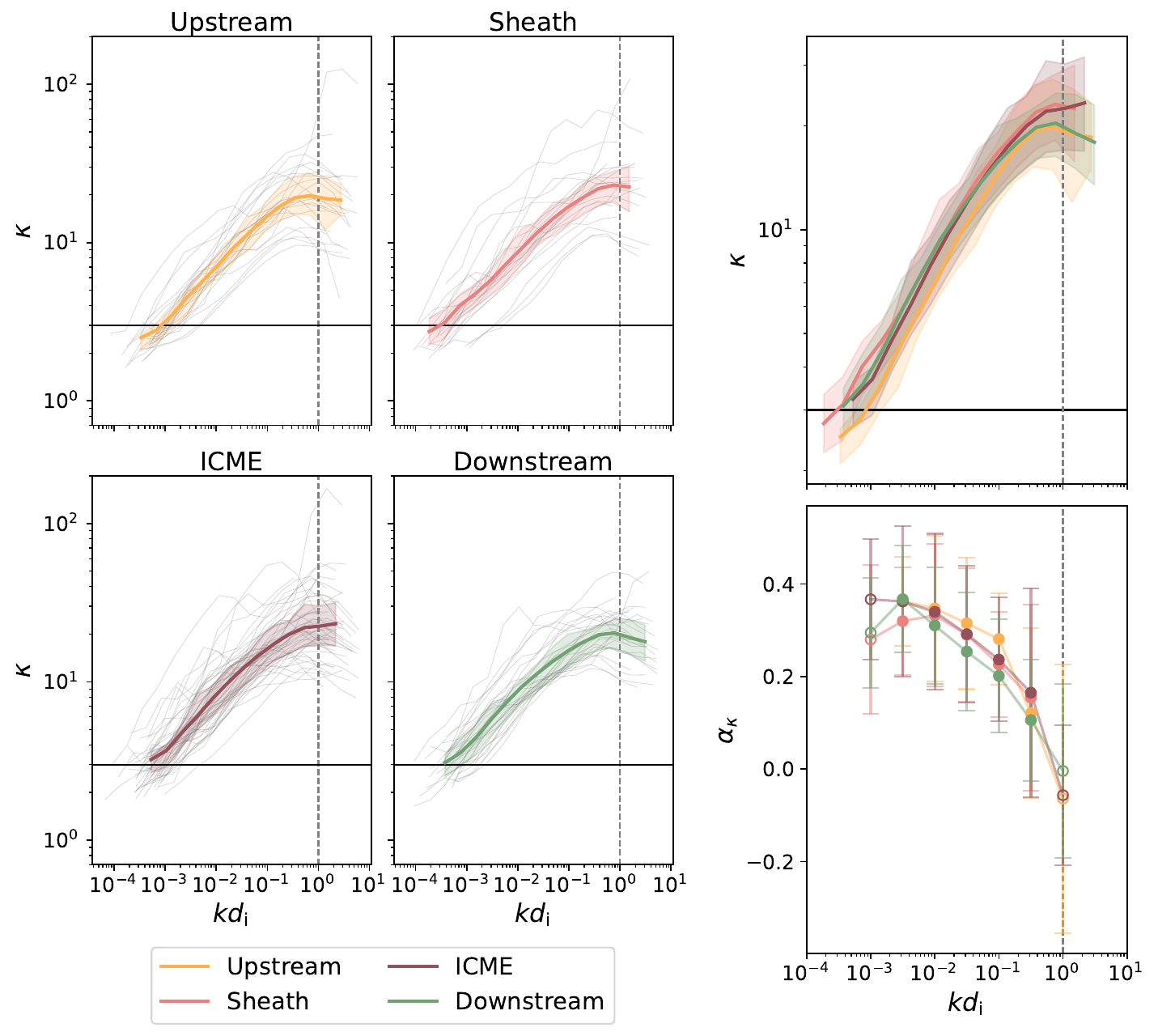}
\caption{Kurtosis as a function of $kd_{\mathrm{i}}$ for all 49 events. Colored line in each panel corresponds to the mean across all events. In the upper right panel means of all four intervals have been plotted with their interquartile ranges. The lower right panel shows the average of event-wise kurtosis scaling exponents at different scales. The bars shown in the panel correspond to the standard deviation of the distributions.
\label{fig3}}
\end{figure*}

The left panels of Figure \ref{fig3} show $\kappa$ of the magnetic field trace as a function of $kd_{\mathrm{i}}$ for different intervals over all events. Individual events are plotted as gray lines and the mean across all events is the colored line in each panel. The means are also plotted in the upper right panel of Figure \ref{fig3} in order to aid comparison between different intervals. The shading in the panels indicates the underlying interquartile range. The largest scale (the smallest $kd_{\mathrm{i}}$) used in the calculation of $\kappa$ was up to 10\% of the interval length to ensure the significance of statistics at large scales \citep[e.g.][]{Song1999}. The validity of the results presented in this section has been checked by removing a running average from the magnetic field data to minimize the effects of the background field. Running averages from 1 to 4 hours were tested and no significant differences were seen in the inertial range.

In the lower right panel of Figure \ref{fig3}, the average of $\alpha_\kappa$ for each event is plotted as a function of $kd_{\mathrm{i}}$. Power-law fitting has been performed in bins across one decade of $kd_{\mathrm{i}}$ values, and the value shown is positioned at the center of each bin. Altogether, the fitting has been done in seven bins but the two outermost bins (empty circles) have been excluded from the rest of the analysis since these bins already include parts of the injection and kinetic scales. The bars in the panel indicate the standard deviation of the distributions to show the spread of variability between the events.

From Figure \ref{fig3} upper right panel, we see that all intervals have a very similar trend and the differences are mostly seen in the behavior of $\kappa$ at the smallest and largest scales. At small scales, the highest $\kappa$ are recorded for the ICMEs and sheaths and the lowest for the upstream and downstream intervals. $\kappa$ for all intervals flattens or peaks at small scales near $kd_{\mathrm{i}} = 1$ indicating a transition to kinetic scales. In the inertial range, $\kappa$ is very similar between all intervals except the upstream, where $\kappa$ starts to decrease faster with increasing scale and reaches the Gaussian level first. Other intervals would reach the Gaussian level around the same $kd_{\mathrm{i}}$ value. The inertial range is the narrowest in the upstream intervals, as defined in terms of the spectral width between $\kappa = 3$ and the $\kappa$ maximum.

From the lower right panel of Figure \ref{fig3} we see that $\alpha_\kappa$ has a decreasing trend when going towards smaller scales, showing that kurtosis increases the fastest at the large-scale end of the inertial range and starts to flatten closer to the kinetic range. This reflects the behavior already seen in the upper right panel in Figure \ref{fig3}. Across most of the inertial range, the downstream interval has slightly lower scaling exponents, indicating a lower level of intermittency among the intervals studied. The sheath and ICME intervals have similar values apart from the largest scales, while the upstream shows higher values around the middle of the inertial range. As shown by the large standard deviations and the example event in Figures \ref{fig1} and \ref{fig2}, where the most intermittent interval was the ICME, there can be significant variation from event to event as to which interval is the most intermittent.

We do not observe a region where the kurtosis scaling exponent remains constant as would be strictly expected of a power-law scaling describing an idealized MHD cascade. Instead, $\alpha_\kappa$ shows some variation across the inertial range. For the rest of the analysis, we concentrate on one range of scales, namely $10^{-2.5}  < kd_{\mathrm{i}} < 10^{-1.5}$, which is located approximately in the middle of the inertial range.

\begin{table}[ht!]
\caption{Kendall correlation coefficients between $\alpha_{\kappa}$ and different parameters in range $10^{-2.5}  < kd_{\mathrm{i}} < 10^{-1.5}$.}
\centering
\begin{tabular}{ccccc}
\hline
             & Upstream & Sheath  & ICME  & Downstream \\ \hline
R            & 0.30   & 0.30  & 0.11 & 0.17     \\
V        & 0.35    & -0.03 & 0.03 & -0.01    \\ 
$\alpha_{\mathrm{PSD}}$     & 0.06   & 0.17  & 0.30 & 0.13     \\
$\sigma_{\mathrm{r}}$     & -0.20    & 0.25  & 0.01  & -0.11    \\
$|\sigma_{\mathrm{c}}|$     & 0.19  & -0.06 & 0.07 & 0.12      \\
$\beta$ & 0.02   & 0.18  & 0.03 & 0.10     \\ \hline
\end{tabular}
\label{table1}
\end{table}

To study the connections between $\alpha_\kappa$ and various parameters, we calculate the Kendall correlation \citep{Kendall1938} between $\alpha_{\kappa}$ and heliocentric distance R, bulk speed V, spectral index $\alpha_{\mathrm{PSD}}$, normalized residual energy $\sigma_{\mathrm{r}}$, absolute value of normalized cross helicity $|\sigma_{\mathrm{c}}|$, and proton $\beta$. The power law fits to $\kappa$ and $S^2_B$ used to determine $\alpha_{\kappa}$ and $\alpha_{\mathrm{PSD}}$, respectively, are made in the range $10^{-2.5}  < kd_{\mathrm{i}} < 10^{-1.5}$. This range is chosen to sit in the middle of the MHD inertial range, away from injection (i.e. large) scales and the transition-kinetic range. To avoid the noise level of the lower-resolution plasma data, the mean $\sigma_{\mathrm{r}}$ and $\sigma_{\mathrm{c}}$ are calculated in the range $10^{-3}  < kd_{\mathrm{i}} < 10^{-2}$. The resulting correlation coefficients are presented in Table \ref{table1}, and in general, all observed Kendall correlation coefficients are low. Correlation coefficients were obtained using other $kd_{\mathrm{i}}$ ranges within the inertial range as well (not shown here). While the values and the parameter with the highest correlation coefficient changed, the values remain low in other $kd_{\mathrm{i}}$ ranges.

Although generally low, some values in Table \ref{table1} suggest some weak dependencies between $\alpha_{\kappa}$ and the parameters studies. We see that in the ICME the highest correlation with $\alpha_{\kappa}$ is recorded with $\alpha_{\mathrm{PSD}}$ and the lowest correlation with $\sigma_{\mathrm{r}}$. The dependence with R is highest in the sheath and upstream while lowest in the ICME. In addition to the correlation with $\alpha_{\mathrm{PSD}}$ in the ICME, the highest correlation values are recorded for the upstream with R and V and the sheath with R and $\sigma_{\mathrm{r}}$.

\begin{figure*}[ht!]
\centering
\includegraphics[width=\linewidth]{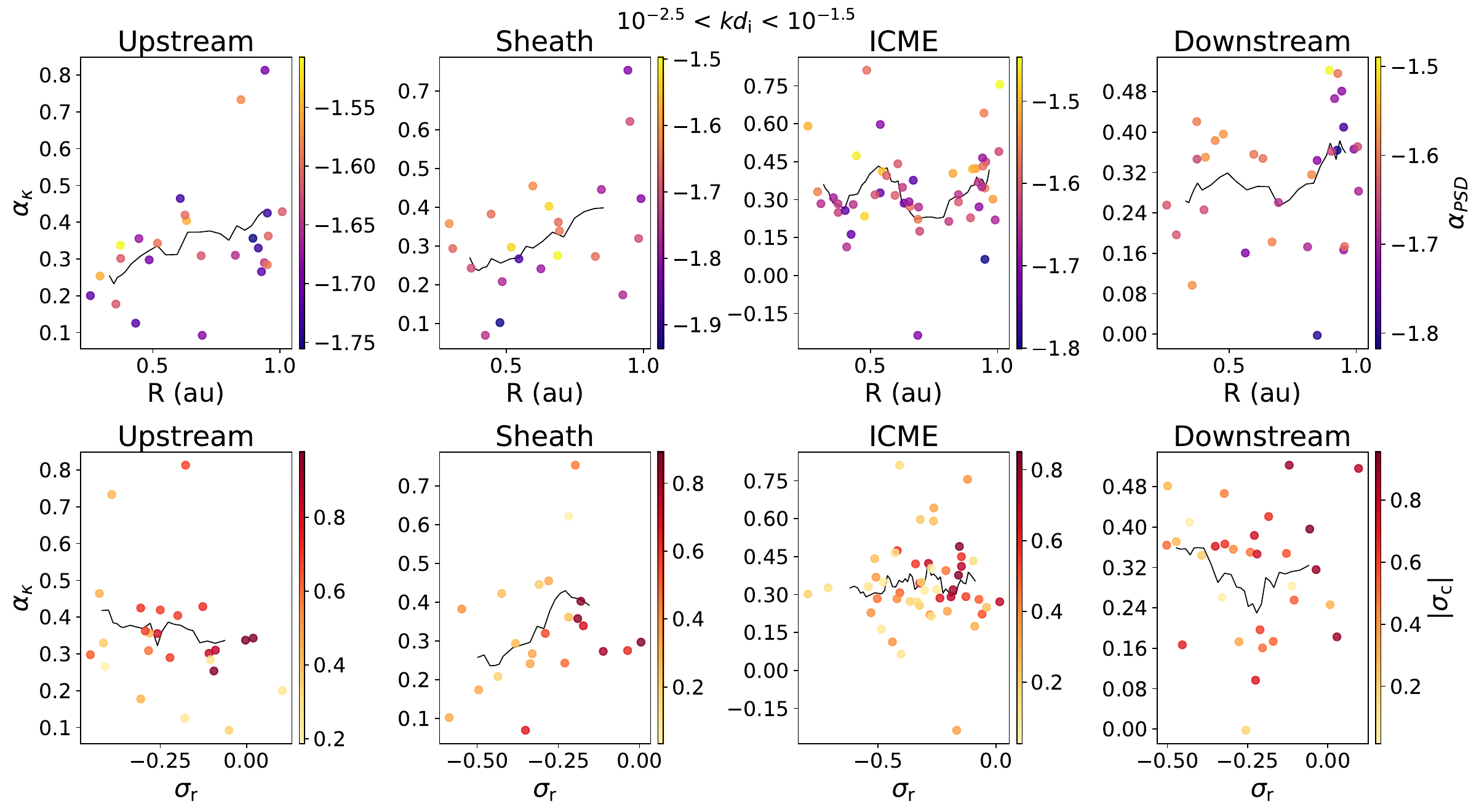}
\caption{$\alpha_\kappa$ in range $10^{-2.5}  < kd_{\mathrm{i}} < 10^{-1.5}$ as a function of R (top row) and $\sigma_{\mathrm{r}}$ (bottom row) for each interval type. The events have been colored based on $\alpha_{\mathrm{PSD}}$ and $\sigma_{\mathrm{c}}$ for the top and bottom panels, respectively. Black lines show five-point moving averages.
\label{fig4}}
\end{figure*}

Figure \ref{fig4} shows further the connection between $\alpha_\kappa$ and R, $\alpha_{\mathrm{PSD}}$, $\sigma_{\mathrm{r}}$, and $|\sigma_{\mathrm{c}}|$. In the top row of the figure, $\alpha_\kappa$ is plotted as a function of R with individual events colored based on their  $\alpha_{\mathrm{PSD}}$ for each interval type. These panels highlight the fact that the evolution of $\alpha_\kappa$ with distance is only seen in the upstream and sheath. Furthermore, the panels show that $\alpha_\kappa$ is the highest with the smallest $|\alpha_{\mathrm{PSD}}|$ values in the sheath and ICME, reflecting the results of the Kendall correlation values.

In the bottom row of Figure \ref{fig4}, $\alpha_\kappa$ is plotted as a function of $\sigma_{\mathrm{r}}$ with individual events colored according to $|\sigma_{\mathrm{c}}|$ for each interval type. Again, the only intervals where an evolution of $\alpha_\kappa$ as a function of $\sigma_{\mathrm{r}}$ is seen are the upstream and the sheath; the trend line shows $\alpha_{\kappa}$ decreasing for the upstream and increasing for the sheath as $\sigma_{\mathrm{r}}$ tends to zero. The connection between $\alpha_\kappa$ and $|\sigma_{\mathrm{c}}|$ is not as clear (reflecting the low correlation values in Table \ref{table1}) apart from the upstream where the lowest $\alpha_\kappa$ are with the lowest $|\sigma_{\mathrm{c}}|$. The expected evolution between $\sigma_{\mathrm{r}}$ and $|\sigma_{\mathrm{c}}|$ is observed, i.e. low $\sigma_{\mathrm{r}}$ values are associated with high $|\sigma_{\mathrm{c}}|$ values.

\section{Discussion} \label{sec:discussion}

While ICMEs have been shown to contain fluctuations and structures at various scales, intermittency in ICMEs is still a relatively new research topic. Due to their well-defined flux rope structures and generally more coherent and smoothly varying nature in the ideal case, one may infer that ICMEs are less intermittent structures than ICME sheaths or the solar wind. Based on the results presented in Figure \ref{fig3}, intermittency behaves similarly in ICMEs as it does in ICME sheaths, and in the solar wind preceding and trailing the ICMEs. When considering intermittency, the exact amplitude of a fluctuation is less important than how the amplitude compares to the amplitudes of other fluctuations at the same scale. So, while ICMEs usually have smaller fluctuation amplitudes than, for example, ICME sheaths, fluctuations may still be intermittent and not conform to a Gaussian distribution, as seen in Figure \ref{fig2}.

Our results for the magnetic field kurtosis are in accordance with the results in \citet{Sorriso-Valvo2021} and \citet{Marquez2023}. In these two case studies, the kurtosis of the ICMEs behaved similarly as it did in the other analyzed intervals and their observed scaling exponents were of the same order as the values reported in this study. Based on the correlation analysis presented in Section \ref{sec:statistical}, there are differences in the dependence between kurtosis and different plasma turbulence parameters in different interval types, but this is not reflected in the general behavior of kurtosis. These differences could indicate different sources of intermittency, including the presence of structures that are not part of the Alfv\'enic turbulence or convective magnetic structures \citep[e.g.][]{Bruno2001, Borovsky2008}.

As shown in Table \ref{table1} and Figure \ref{fig4}, no radial evolution of intermittency is seen in the ICME or downstream. In this sense ICMEs are similar to the slow solar wind, where turbulence appears to be in a fully developed state and no radial evolution of intermittency is seen \citep{Bruno2003}. It may be in the case of ICMEs that, while they expand and interact with the solar wind, they are relatively stable, static structures that retain their small-scale statistical properties during their propagation from the Sun to 1 au. Related indications of the static nature of ICMEs in the inertial range have previously been observed in \citet{Good2023}. In the sheath, intermittency increases with radial distance. A possible explanation is that the sheaths are still evolving structures with new plasma being added in as the ICME propagates. The sheaths have most likely not reached a state of fully developed turbulence at the smallest radial distances examined in this study and thus, increasing intermittency with distance is observed. The same is also likely true for the upstream intervals.

\begin{figure}[ht!]
\centering
\includegraphics[width=\linewidth]{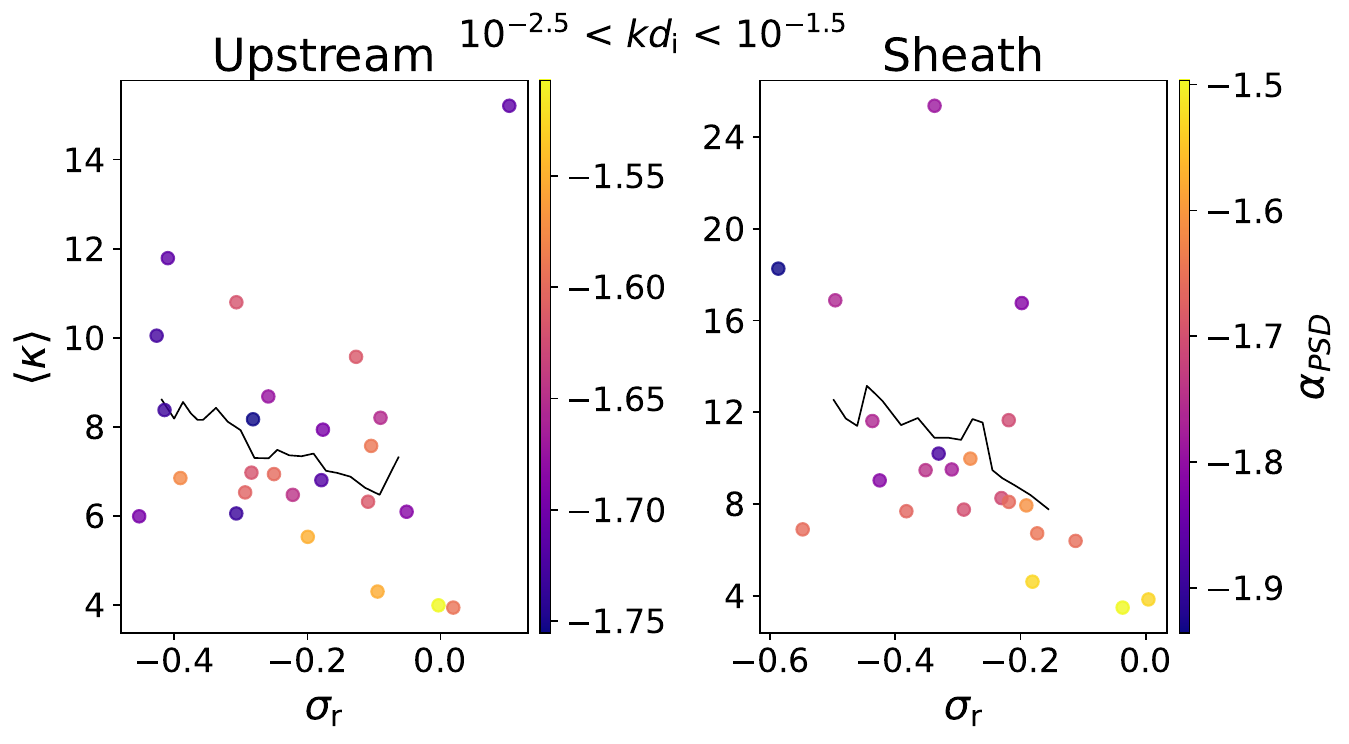}
\caption{$\langle \kappa \rangle$ in range $10^{-2.5}  < kd_{\mathrm{i}} < 10^{-1.5}$ as a function of $\sigma_{\mathrm{r}}$ for the sheath and ICME intervals. The events have been colored based on $\alpha_{\mathrm{PSD}}$. Black lines show five-point moving averages.}
\label{fig5}
\end{figure}

The bottom row of Figure \ref{fig4} showed no clear evolution of $\alpha_{\kappa}$ with $\sigma_{\mathrm{r}}$ or $|\sigma_{\mathrm{c}}|$ in ICMEs and downstream intervals. In the upstream, we observe a slightly decreasing trend in $\alpha_{\kappa}$ as $\sigma_{\mathrm{r}}$ becomes less negative, while in the sheath there is the opposite trend with $\sigma_{\mathrm{r}}$: $\alpha_{\kappa}$ increases as $\sigma_{\mathrm{r}}$ becomes less negative. Thus, the events with the lowest $\alpha_{\kappa}$ (i.e. the least developed turbulent cascade) are associated with the most negative residual energy. These events are also associated with the steepest spectral slopes. To make our results more comparable with other studies, in Figure \ref{fig5} we replot the upstream and sheath panels from the bottom row of Figure \ref{fig4} with the mean kurtosis value $\langle \kappa \rangle$ instead of $\alpha_{\kappa}$, calculated in the same $kd_{\mathrm{i}}$ range. Here, the events are colored according to their $\alpha_{\mathrm{PSD}}$ values. The ICME and downstream intervals are omitted since no evolution with $\sigma_{\mathrm{r}}$ is observed. For the downstream this result is similar to those from MHD simulations of the solar wind by \citet{Shi2025}. In the upstream and the sheath, $\langle \kappa \rangle$ is seen to increase as $\sigma_{\mathrm{r}}$ becomes more negative. A similar trend was observed by \citet{Bowen2018} in the solar wind. The trend of $\alpha_{\kappa}$ and $\langle \kappa \rangle$ with $\sigma_{\mathrm{r}}$ is the same in the upstream, but in the case of the sheath, the trend is opposite for $\alpha_{\kappa}$ and $\langle \kappa \rangle$ as $\sigma_{\mathrm{r}}$ becomes more negative. It may be the case that, in the sheaths with strongly negative $\sigma_{\mathrm{r}}$ and high $\langle \kappa \rangle$ but low $\alpha_{\kappa}$, the high $\langle \kappa \rangle$ could be due to the significant presence of fluctuations and coherent structures generated by dynamics other than a very well-developed turbulent cascade; indeed, the turbulence could even be weakly developed (i.e. low $\alpha_{\kappa}$) in high-kurtosis intervals such as these. These structures could also explain some of the very steep spectral slopes seen in Figure \ref{fig5} \citep{Li2011}. If this is the case, the absolute kurtosis values might not be the best measure of intermittency in relation to the MHD turbulence because they do not take into account the scale-to-scale evolution given by $\alpha_\kappa$. The connection between kurtosis values and the change in kurtosis between scales will be the subject of future studies. In addition, the sheaths themselves may prove to be interesting future research subjects. The exact nature of intermittent fluctuations in ICME sheaths as well as their radial evolution would benefit from further studies.

\section{Conclusions} \label{sec:conclusions}

In this paper, we have presented a statistical study of intermittency in 49 ICMEs and their sheath regions, and also upstream and downstream solar wind intervals. The main intermittency measure used in the study was kurtosis, with its scaling exponent $\alpha_\kappa$ used to compare the level of intermittency in different time series. The overall behavior of kurtosis is seen to be similar in all interval types and indicative of the presence of intermittency. The average intermittency level is similar in all studied intervals but the radial evolution of $\alpha_\kappa$ is only seen in the upstream solar wind and ICME sheaths. In ICMEs, the radial invariance of intermittency suggests that they are a relatively stable or fully developed turbulent environment. The sheaths are dynamically younger structures due to the steady input of new solar wind plasma during the propagation of the ICME and thus, turbulence is not yet fully developed and the level of intermittency rises as the distance from the Sun increases. In addition, the analysis of $\alpha_\kappa$ and $\langle \kappa \rangle$ may suggest that, in the sheaths, some of the intermittent fluctuations correspond to non-turbulent structures.

\begin{acknowledgments}

The HELIO4CAST ICMECAT living catalog is available online at \url{https://helioforecast.space/icmecat} and archived at \url{https://doi.org/10.6084/m9.figshare.6356420} \citep{Moestl2020}. For this study, version 21, released 11 April 2024, was used.

This work was funded by a Research Council of Finland Fellowship (grants 338486, 346612, and 359914; INERTUM). E.K., S.G., and J.R. acknowledge the Finnish Centre of Excellence in Research of Sustainable Space (Research Council of Finland grant number 312390). E.K. acknowledges the ERC under the European Union's Horizon 2020 Research and Innovation Programme Project SolMAG 724391. C. M. is supported by ERC grant (HELIO4CAST, 10.3030/101042188) funded by the European Union. Views and opinions expressed are those of the authors only and do not necessarily reflect those of the European Union or the European Research Council Executive Agency. Neither the European Union nor the granting authority can be held responsible for them. We thank the Solar Orbiter and Parker Solar Probe instrument teams for providing the data used in this work.

\end{acknowledgments}

%






\appendix
\label{sec:appendix}

The shock, start and end times of analyzed ICMEs are shown in Table \ref{table2}. The times are taken from HELIO4CAST ICMECAT database and the one event with changed end time has been marked with an asterisk (*). In addition, the table lists $\alpha_\kappa$ values for each event in the range $10^{-2.5}  < kd_{\mathrm{i}} < 10^{-1.5}$.

\begin{table}[ht!]
\caption{List of analyzed ICMEs}
\centering
\begin{tabular}{cccccccccc}
\cline{7-10}
\multicolumn{1}{l}{} & \multicolumn{1}{l}{} & \multicolumn{1}{l}{} & \multicolumn{1}{l}{} & \multicolumn{1}{l}{} & \multicolumn{1}{l}{} & \multicolumn{4}{c}{$\alpha_\kappa$}                                                                       \\ \hline
Event          & Shock Time       & Start Time       & End Time        & Spacecraft           & R                & Upstream  & Sheath               & ICME                 & Downstream \\
\multicolumn{1}{c}{Number} & \multicolumn{1}{c}{(UT)} & \multicolumn{1}{c}{(UT)} & \multicolumn{1}{c}{(UT)} & \multicolumn{1}{c}{} & \multicolumn{1}{c}{(au)} & \multicolumn{1}{c}{} & \multicolumn{1}{c}{} & \multicolumn{1}{c}{} & \multicolumn{1}{c}{}  \\ \hline
1            & ...              & 2018-11-11 23:51 & 2018-11-12 05:59  & PSP        & 0.25   & 0.20                 & ...    & 0.59  & 0.26                  \\
2            & 2019-03-15 09:00 & 2019-03-15 12:11 & 2019-03-15 17:49  & PSP        & 0.55   & ...                 & 0.27   & 0.41  & ...                   \\
3            & ...              & 2020-05-28 08:50 & 2020-05-28 14:59  & PSP        & 0.35   & 0.18                & ...    & 0.31  & 0.10                   \\
4            & ...              & 2021-05-06 18:26 & 2021-05-07 15:07  & SolO       & 0.91   & 0.33                & ...    & 0.42  & 0.47                  \\
5            & 2021-05-10 06:24 & 2021-05-10 14:01 & 2021-05-11 11:39  & SolO       & 0.92   & ...                 & 0.17   & 0.37  & 0.36                  \\
6            & ...              & 2021-05-27 20:14 & 2021-05-28 10:27  & SolO       & 0.95   & 0.42                & ...    & 0.35  & 0.17                  \\
7            & ...              & 2021-08-25 01:06 & 2021-08-25 10:43  & SolO       & 0.63   & 0.40                 & ...    & 0.29  & 0.35                  \\
8            & 2021-10-05 01:50 & 2021-10-05 05:09 & 2021-10-05 11:00  & SolO       & 0.66   & ...                 & 0.40    & 0.27  & ...                   \\
9            & 2021-11-03 14:03 & 2021-11-04 01:25 & 2021-11-04 19:47  & SolO       & 0.85   & 0.73                & 0.45   & 0.29  & 0.00                  \\
10           & 2021-11-27 23:00 & 2021-11-28 11:28 & 2021-11-29 06:50  & SolO       & 0.99   & ...                 & 0.42   & 0.22  & 0.37                  \\
11           & 2022-01-25 13:05 & 2022-01-25 14:02 & 2022-01-27 07:15  & SolO       & 0.89   & 0.36                & ...    & 0.23  & 0.52                  \\
12           & 2022-03-07 22:26 & 2022-03-08 05:51 & 2022-03-08 14:44  & SolO       & 0.49   & 0.30                 & 0.21   & 0.81  & ...                   \\
13           & 2022-03-08 14:46 & 2022-03-08 21:37 & 2022-03-09 06:56  & SolO       & 0.48   & ...                 & 0.10    & 0.23  & 0.40                   \\
14           & 2022-04-08 13:50 & 2022-04-08 20:16 & 2022-04-09 23:59  & SolO       & 0.42   & ...                 & 0.07   & 0.16  & ...                   \\
15           & 2022-05-02 17:18 & 2022-05-02 22:09 & 2022-05-03 04:55  & PSP        & 0.69   & ...                 & 0.28   & 0.22  & ...                   \\
16           & 2022-05-20 04:19 & 2022-05-20 08:09 & 2022-05-20 13:14  & PSP        & 0.45   & 0.36                & 0.38   & 0.47  & 0.38                  \\
17           & ...              & 2022-06-22 18:24 & 2022-06-23 05:07  & SolO       & 1.00    & ...                 & ...    & 0.49  & 0.37                  \\
18           & 2022-07-02 04:42 & 2022-07-02 20:02 & 2022-07-04 19:42  & PSP        & 0.69   & 0.09                & 0.34   & 0.18  & 0.26                  \\
19           & ...              & 2022-07-13 15:55 & 2022-07-14 07:09  & SolO       & 1.01   & 0.43                & ...    & 0.76  & 0.28                  \\
20           & 2022-07-25 06:24 & 2022-07-25 11:45 & 2022-07-26 01:40* & SolO       & 0.98   & ...                 & 0.32   & 0.30   & ...                   \\
21           & 2022-09-06 10:01 & 2022-09-07 06:47 & 2022-09-08 04:10  & SolO       & 0.69   & ...                 & ...    & -0.24 & ...                   \\
22           & 2022-09-08 21:02 & 2022-09-08 23:09 & 2022-09-09 05:38  & SolO       & 0.67   & ...                 & ...    & 0.38  & 0.18                  \\
23           & ...              & 2022-09-18 06:28 & 2022-09-18 12:30  & PSP        & 0.43   & 0.13                & ...    & 0.28  & ...                   \\
24           & ...              & 2023-01-12 11:31 & 2023-01-12 20:11  & SolO       & 0.95   & 0.36                & ...    & 0.45  & 0.17                  \\
25           & 2023-02-17 14:03 & 2023-02-17 23:23 & 2023-02-18 05:18  & SolO       & 0.82   & 0.31                & 0.27   & 0.40   & 0.32                  \\
26           & ...              & 2023-02-26 13:35 & 2023-02-26 19:42  & PSP        & 0.56   & ...                 & ...    & 0.39  & 0.16                  \\
27           & ...              & 2023-03-09 05:58 & 2023-03-09 16:04  & SolO       & 0.65   & ...                 & ...    & 0.29  & ...                   \\
28           & 2023-04-10 04:34 & 2023-04-10 10:12 & 2023-04-10 13:16  & SolO       & 0.29   & 0.25                & 0.36   & 0.33  & 0.20                   \\
29           & 2023-05-03 10:24 & 2023-05-03 10:34 & 2023-05-03 16:06  & SolO       & 0.54   & ...                 & ...    & 0.60   & ...                   \\
30           & ...              & 2023-06-11 07:02 & 2023-06-11 15:27  & PSP        & 0.41   & ...                 & ...    & 0.11  & 0.35                  \\
31           & 2023-06-26 00:16 & 2023-06-26 23:27 & 2023-06-27 20:16  & SolO       & 0.94   & 0.81                & ...    & 0.46  & ...                   \\
32           & 2023-06-27 20:17 & 2023-06-27 23:28 & 2023-06-28 06:21  & SolO       & 0.94   & ...                 & 0.75   & 0.43  & 0.48                  \\
33           & ...              & 2023-07-01 15:14 & 2023-07-02 06:37  & PSP        & 0.37   & 0.30                 & ...    & 0.25  & 0.35                  \\
34           & 2023-07-18 17:07 & 2023-07-18 18:40 & 2023-07-19 02:23  & SolO       & 0.95   & ...                 & ...    & 0.64  & ...                   \\
35           & ...              & 2023-07-22 20:40 & 2023-07-23 02:10  & SolO       & 0.94   & 0.29                & ...    & 0.35  & ...                   \\
36           & 2023-07-26 22:00 & 2023-07-27 00:58 & 2023-07-27 04:20  & SolO       & 0.93   & 0.27                & ...    & 0.27  & 0.52                  \\
37           & ...              & 2023-08-18 23:11 & 2023-08-19 10:15  & SolO       & 0.81   & ...                 & ...    & 0.21  & 0.17                  \\
38           & 2023-09-06 20:23 & 2023-09-07 06:48 & 2023-09-07 17:41  & SolO       & 0.63   & 0.42                & 0.24   & 0.35  & ...                   \\
39           & 2023-09-08 19:32 & 2023-09-08 20:57 & 2023-09-09 06:42  & SolO       & 0.61   & 0.46                & ...    & 0.44  & ...                   \\
40           & 2023-09-09 10:41 & 2023-09-09 18:41 & 2023-09-10 01:20  & SolO       & 0.60    & ...                 & 0.45   & 0.32  & 0.36                  \\
41           & 2023-09-10 10:05 & 2023-09-10 12:00 & 2023-09-10 20:00  & PSP        & 0.54   & ...                 & ...    & 0.33  & ...                   \\
42           & ...              & 2023-09-17 13:52 & 2023-09-17 18:00  & PSP        & 0.40    & ...                 & ...    & 0.26  & 0.25                  \\
43           & 2023-10-10 22:32 & 2023-10-11 03:27 & 2023-10-11 09:19  & SolO       & 0.31   & ...                 & 0.29   & 0.28  & ...                   \\
44           & 2023-10-17 10:24 & 2023-10-17 17:46 & 2023-10-18 11:50  & SolO       & 0.37   & 0.34                & 0.24   & 0.28  & 0.42                  \\
45           & 2023-10-28 06:06 & 2023-10-28 14:11 & 2023-10-28 20:54  & SolO       & 0.52   & 0.34                & 0.30    & 0.32  & ...                   \\
46           & 2023-11-11 09:14 & 2023-11-12 03:57 & 2023-11-12 20:21  & SolO       & 0.69   & 0.31                & 0.36   & 0.27  & ...                   \\
47           & 2023-12-01 02:27 & 2023-12-01 03:11 & 2023-12-01 10:27  & SolO       & 0.85   & ...                 & ...    & 0.32  & 0.34                  \\
48           & ...              & 2023-12-11 16:54 & 2023-12-12 00:32  & SolO       & 0.90    & ...                 & ...    & 0.42  & 0.36                  \\
49           & 2023-12-29 02:28 & 2023-12-29 11:30 & 2023-12-29 14:45  & SolO       & 0.95   & 0.28                & 0.62   & 0.06  & 0.41                                   \\ \hline
\end{tabular}
\label{table2}
\end{table}


\bibliography{bibliography}{}
\bibliographystyle{aasjournal}



\end{document}